# Impurity effects on binding energy, diamagnetic susceptibility and photoionization cross-section of chalcopyrite AgInSe2 Nanotadpole

Grigor A. Mantashian[1], David B. Hayrapetyan[1]

[1] Department of General Physics and Quantum Nanostructures, Russian-Armenian University, Yerevan, Armenia

E-mail: grigor.mantashyan@rau.am

## Abstract

Recently the interest in chalcopyrite semiconductor nanostructures has increased because of their non-toxicity and their wide direct bandgap. Likewise, structures with non-trivial geometry are particularly interesting because of their electronic, optical, and magnetic properties. In the current article, the finite element method was used in conjunction with the effective mass approximation to theoretically investigate the properties of a chalcopyrite $AgInSe_2$ nanotadpole in the presence of an off-center impurity. The morphology of the nanotadpole gives it excellent hydrodynamic properties and is ideal for a wide range of applications. The probability densities for various impurity positions and energy levels were obtained. The results suggested a strong dependence of the behavior of the electron on the impurity positions and the orientation of the wave function. The investigation of the nanotadpole's energy spectra and their comparison with the cylindrical and spherical quantum dots suggest that the spectrum has degenerate states similar to the spherical case, however at some ranges, the levels behave similarly to the cylindrical case. The binding energy's dependence on the nanotadpole's size and the impurity position was obtained. The dependence of the diamagnetic susceptibility on the impurity position was calculated. An extensive investigation of the photoionization cross-section was carried out for the ground and the first two excited states as the initial states and the first twenty excited states as the final states.

Keywords: nanotadpole, chalcopyrite $AgInSe_2$, quantum dots, binding energy, diamagnetic susceptibility, photoionization cross-section, impurity

## 1. Introduction

The past decades were particularly fruitful for the field of semiconductor nanostructures, quantum dots (QDs) however garnered a special interest. Their unique tunable optical properties have led them to find various applications such as biological markers, light-emitting devices, solar cells, advanced batteries, etc. [1-4]. Despite that, the unfortunate fact is that the best-developed semiconductor QDs have a serious problem, they contain highly toxic elements and for that reason, they have limited commercial applications. A way out of this predicament can be found by placing our focus on multinary semiconductor QDs, which consist of I-III-VI2 semiconductors and their solid solutions, for example, $AgInSe_2$, $AgInS_2$, $ZnS–AgInS_2$, $CuInS_2$, and $CuInS_2–CuGaS_2$ [5-9]. Besides consisting of non-toxic semiconductor materials, these QDs have a direct bandgap, have high photoluminescence quantum yield in visible near-infrared regions, and have a large absorption coefficient, they also exhibit interesting magnetic, thermoelectric, and thermal properties [10-12].

Within the group of I-III-VI2 compounds, AgInSe2 is a promising material because of its direct bandgap energy of 1.2 eV and a wide array of applications such as solar cells, bio-imaging, diamond-like thermoelectric materials, photocatalysis, and biosensors [13-19].

Moreover, there have been considerable successes in the synthesis of I-III-VI2 QDs with non-trivial geometries [20-22], which have shown interesting optical, electronic, and magnetic properties, as well as having wider applications compared to the nanostructures with more common shapes. In particular, QDs with tadpole-like morphology i.e. structures with a clearly defined head and tail. AgInSe2 nano tadpoles (NT) that are investigated in the current article are obtained from a single molecular precursor by the authors of the following article [23]. It was found that AgInSe2 NTs possess even greater saturable absorption properties than similarly sized nanorods significant third-order nonlinear properties were observed. These structures are useful because of their natural good hydrodynamic properties, which allows them to be particularly useful in drug delivery and bioimaging; that is why non-toxic NTs have special importance. It is also important to note that because of their asymmetric morphology the AgInSe2 NTs exhibit drastically different optical properties for different incident light angles. The complex geometry of these structures makes the analytical solution of the discussed practically impossible, leaving either approximate methods that will give imprecise solutions or numerical methods. As such the finite element method (FEM) was used to solve the problem.

The electronic structure in the presence of impurity for trivial geometries is a widely investigated problem [24-27]. For example, the authors of the following article [25] have studied the effects of impurities and external fields on nonlinear optical rectification. The diamagnetic susceptibility is one of the most interesting magnetic properties of a QD with a hydrogenic impurity, as such, it is a widely studied topic [28-31]. The authors of the following work have investigated the diamagnetic susceptibility and binding energies in GaN/AlGaN QD's with and without the Zeeman effect [29]. It was found that the binding energy is higher when the Zeeman effect is present.

We have considered a system with an off-center hydrogenic impurity. One of the most important properties that can be calculated for such a system is diamagnetic susceptibility (DS). The effects of the hydrostatic pressure on DS of an on-center hydrogenic donor impurity in core/shell/shell structure with Kratzer confining potential has been theoretically investigated by the authors of this work [32]. As the next logical step, we have calculated the binding energy, the photoionization cross section's (PCS), dependence on different parameters. These properties were thoroughly investigated in various works such as [33] among others.

Because of their special properties and their relative novelty, the theoretical investigation of AgInSe2 NTs is an important problem, which has not been addressed before. Unlike structures that were thoroughly investigated the investigation of the NT's electronic wavefunctions and the energy spectrum is an unaddressed problem. The paper is structured as follows: the morphological and material properties together with the impurity problem are presented in section 2.1. The theoretical formulas of the investigated physical properties are presented in section 2.2. The behavior of the probability density and the energy spectrum is discussed in section 3.1. The DS's dependence was thoroughly discussed in section 3.2. Finally, the results for the Photoionization Cross Section's dependence on the incident light were discussed in section 3.3. The matrix element for the discussed transitions is presented in the Appendix 1.

## 2. Materials and Methods

NT's have neither a spherical nor cylindrical symmetry, instead, they have a non-trivial hybrid symmetry. That fact makes the eigenproblem easily solvable in the cartesian coordinates. Most structures with non-trivial geometry are impossible to investigate analytically. Sometimes analytical approximation methods can be employed, however, that leads to the diminishing of the solution's precision. Taking all of the aforementioned facts into account the FEM was used in conjunction with the effective mass approximation to solve this problem.

### 2.1 Material and Morphological Properties

It was decided that the head of the tadpole can be approximated with a sphere, with the radius $R_{head}$. The tail of the NT was approximated by a cylinder with the length of $L_{tail}$ and the radius $R_{tail}$. The impurity position was chosen to be centered with the respect to the z-axis $\vec{r}_{imp}(0,0,z_{imp})$, because of that throughout the article the $z_{imp}$ is referred to as the position of the impurity. The vertical cross-section of the NT is presented in Figure 1.

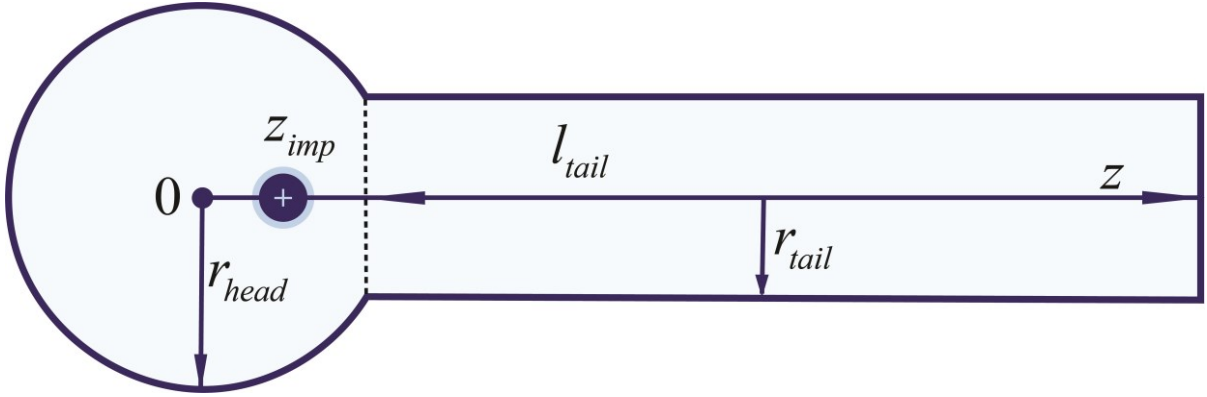

**Figure 1**. *The schematic representation of the chalcopyrite AgInSe$_2$ nanotadpole with the impurity centered with respect to the z-axis. The morphological parameters are as follows: $R_{head}$ - is the radius of the head, $R_{tail}$ - is the radius of the tail, $L_{tail}$ - is the length of the tail, $z_{imp}$ - is the position of the impurity*

The next obvious step is choosing the material, a chalcopyrite AgInSe2 was chosen because the structure was experimentally realized has a wide direct bandgap and finally because of its non-toxicity. The material parameters that were used in the calculation are as follows: dielectric permeability - $\varepsilon = 10.73$, infrared refractive index $n = 2.82$, electron's effective mass- $m^* = 0.12 \cdot m_0$, Bohr radius- $a_b = 4.87$, Rydberg's energy- $E_R = 13.44 meV$, system charge density divided by unit volume - $\sigma = \frac{n}{V} = \frac{3n}{4\pi a^2 c} \approx 4.43 \cdot 10^6 m^{-3}$. The data was taken from [34], [35]. And finally, the Hamiltonian operator representing the impurity states in such a system can be written as:

$$\hat{H} = -\frac{\hbar^2}{2m^*}\Delta - \frac{e^2}{\varepsilon|\vec{r}-\vec{r}_{imp}|} + U_{conf}(\vec{r}) \quad (1)$$

where $e$ -is the electron's charge, $r$ - a distance of the electron from impurity; $U_{conf}(\vec{r})$ - is the confinement potential of the following form:

$$U_{conf}(\vec{r}) = \begin{cases} U_{conf}(inside) = 0 \\ U_{conf}(outside) = \infty \end{cases} \quad (2)$$

As we mentioned above the impurity is centered with the respect to the z-axis $\vec{r}_{imp}(0,0,z_{imp})$ and from this point on is referred to as $z_{imp}$.

The FEM calculation consists of the following steps: definition of the original continuous region in which the problem is solved, the definition of the partial differential equation (PDE), the definition of the boundary conditions, discretization of the mesh into a finite number of elements, systematically recombining all element equations into a system of equations for the final calculation. To complete the FEM calculation, we apply the Hamiltonian operator to the three-dimensional wavefunction $\Psi(x,y,z)$, the obtained PDE is solved in the confines of the mesh the vertical cross-section of which is given in Figure 1, while assuming that the probability density turns to zero at the structure's boundaries. This assumption is justified by the fact that in the experiment [23] the NT's are dissolved in a dielectric solvent.

*2.2 Theory*

In the next part of this section, we will proceed to present the formulas used to calculate the physical properties. The DS has been calculated using the formula taken from [36]:

$$\chi_{dia} = -\frac{e^2}{6m^*c^2}\int_0^\infty \left\{\Psi_{n,l,m}^0(\vec{r})\right\}^2 r^4 dr \quad (3)$$

Where $c$ - is the speed of light. The PCS dependent on the incident photon energy was calculated with the formula taken from [37]:

$$\sigma(\hbar\omega) = \left[\left(\frac{F_{eff}}{F_0}\right)^2 \frac{n}{k}\right]\frac{4\pi^2}{3}a_{FS}\hbar\omega\sum_f M_{if}\delta(E_f - E_i - \hbar\omega) \quad (4)$$

Where the $\delta(E_f - E_i - \hbar\omega)$ function is replaced by a narrow Lorentzian, which has the following form:

$$\delta(E_f - E_i - \hbar\omega) = \frac{\hbar\Gamma}{\pi\left(\left(\hbar\omega - (E_f - E_i)\right)^2 + (\hbar\Gamma)^2\right)} \quad (5)$$

In the absence of experimental data, the $\Gamma$ is taken to be $\Gamma = 0.1 meV$. And $M_{if} = \left|\langle \Psi_i | \vec{er} | \Psi_f \rangle\right|^2$ - is the squared dipole matrix element of the optical transition, describes the probability of transitions.

Here $\vec{e}$ - is the polarization vector. As the initial state for the summation of the PCS the ground state and the first two excited states $i = 1, 2, 3$ were chosen. The first twenty excited states were chosen as the final states for the PCS $f = 2, 3 ..., 21$.

## 3. Results and Discussion

Because of the novelty of the chalcopyrite AgInSe2 NT's and the lack of theoretical investigation, it is important to focus on the behavior of the probability density (PD) and the energy for this structure.

### *3.1 Probability Density and Energy*

The results that were obtained suggested that eigenstates and eigenvalues in the structure behave neither as a spherical QD nor as a cylindrical QD. Thus, the numeration of the quantum states cannot be done in the analogy with these structures. Because of that, the states are numerated with the number $n = 1, 2, 3 .....$. In our calculations, unless stated otherwise the morphological parameters of the chalcopyrite AgInSe2 NT's are taken to be the following: the head diameter $R_{head} = 60 nm$, tail diameter $R_{tail} = 30 nm$, and tail length $L_{tail} = 240 nm$. These are the average sizes of TPs taken from the experimental work [23]. It is important to note that NT's with different morphologies i.e. different head to tail ratios have been examined and there are three distinct cases. In the first case where $R_{head} = 60 nm$ and $20 nm \leq R_{tail} < 25 nm$ the difference between the confinement is large as a result, the PD is entirely localized in the head for the smaller NT's in that range, and the NT's on the higher end of that range show small visible "seep through". The impurity positions outside of the head region for that morphology have no practical influence on the behavior of the PD. And the electron behaves analogously to the spherical QD. In the second case where $R_{head} = 60 nm$ and $35 nm < R_{tail} < 55 nm$ the confinement effects in the head and tail are fairly close as a result, the electron is localized in the NT relatively evenly without the impurity. As a result, the impurity position has a very large contribution. In this case, the PD behaves analogously to the cylindrical QD. The final case is the "hybrid" case where $R_{head} = 60 nm$ and $25 nm < R_{tail} < 35 nm$ this range is the closest to the experimental values and presents the most interesting physical properties. The intermediate state is thoroughly investigated in the current article.

The density plot of the PD for the electron in the ground state $n = 1$ for the following positions: $z_{imp} = 0 nm, 50 nm, 100 nm, 150 nm$ of the impurities is presented in Figure 2. The figure clearly demonstrates that for the ground state the evolution of the impurity positions has a small visible impact on the wave function which is mainly localized in the head region. The two cases worth discussing are $z_{imp}=0nm, 150nm$ : in the $z_{imp}=0$ case the impurity traps the PD in the head region, and when $z_{imp}=150nm$ the impurity is too far to affect the PD, a small amount of "seep through" can be seen.

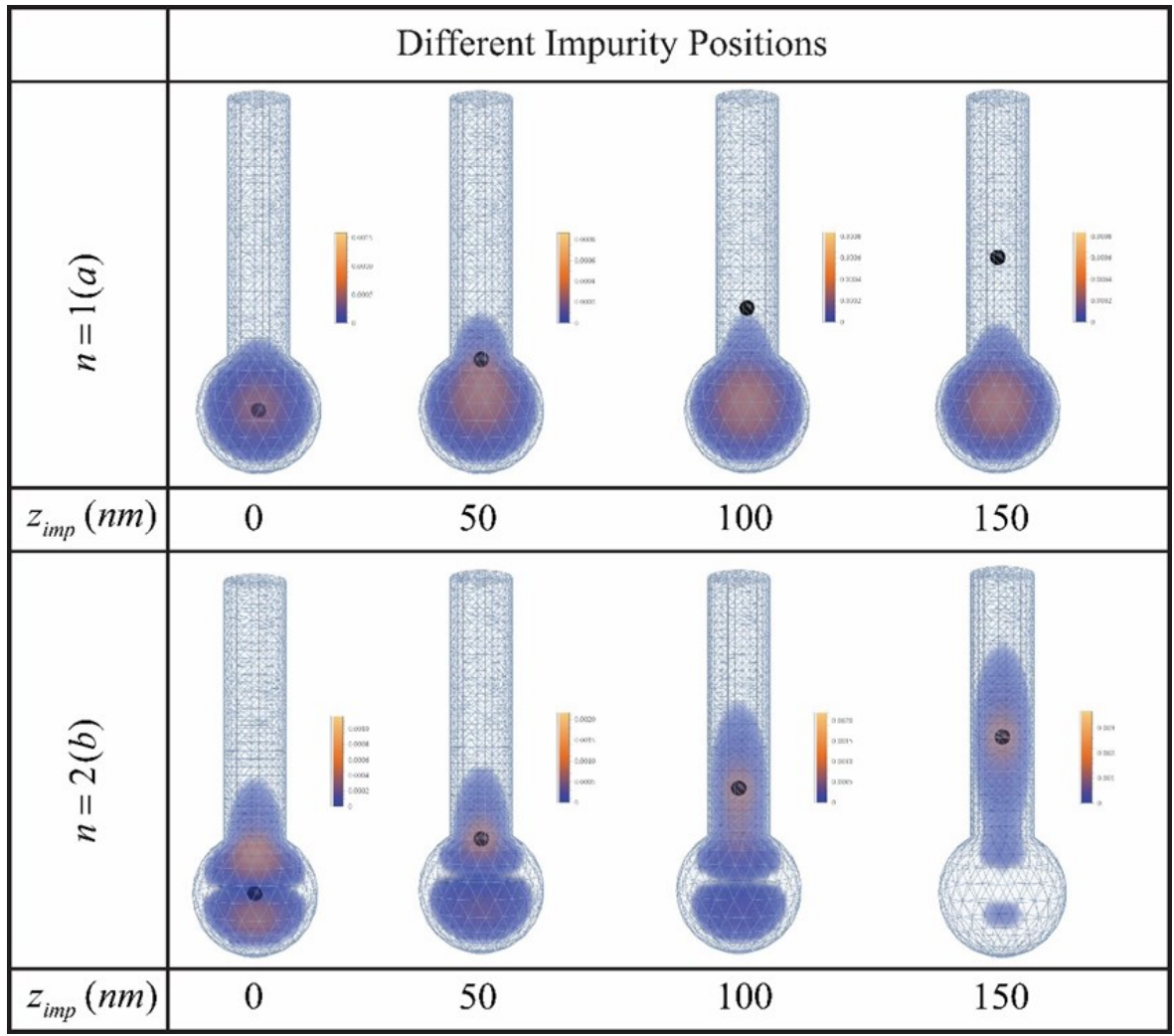

**Figure 2.** *The plots of the probability density for the $n=1$ and $n=2$ with four different impurity positions for chalcopyrite AgInSe$_2$ nanotadpole are presented on the left. The figure suggests that for the $n=1$ case the probability density behaves analogous to the sphere, the position of the impurity position (which is presented by the black dot) $z_{imp}$ has little bearing on the behavior. For the $n=2$ picture changes with the impurity position $z_{imp}$ having a significant contribution to the form of the probability density, in some cases, it behaves like the cylindrical case.*

Contrasting the aforementioned, the PD of the electron in the first excited state $n=2$ clearly shows that the impurity position has a greater impact on the PD of the electron. Let us discuss the four presented cases:

- $z_{imp}=0$ the electron is primarily localized in the head region with some small probability of the electron being in the tail;

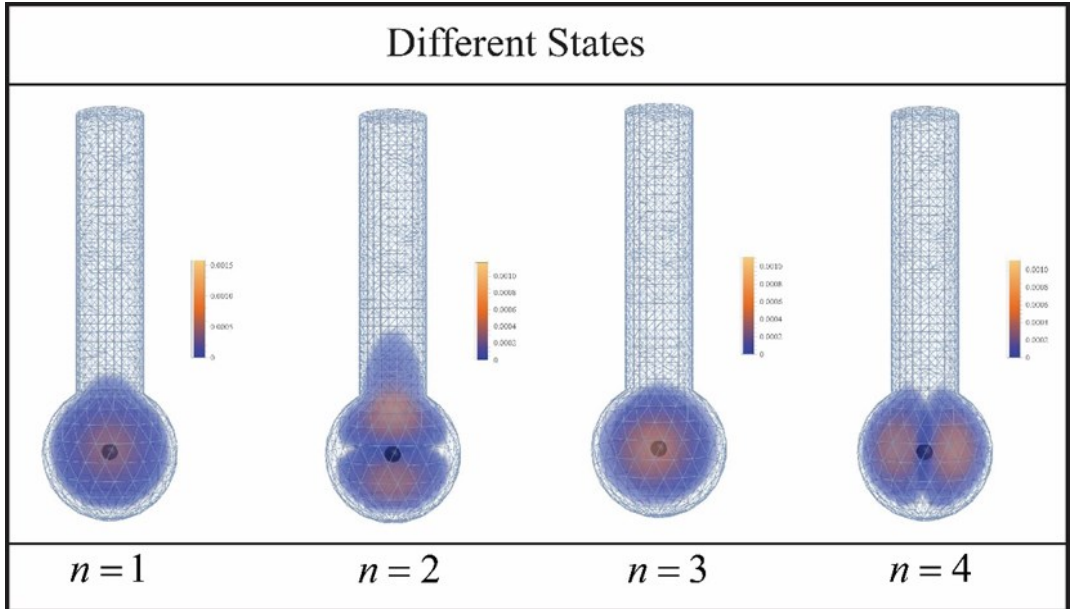

**Figure 3.** *The density plots of the first four states with the impurity centered in the head $z_{imp} = 0$ for the chalcopyrite AgInSe$_2$ nanotadpole are presented. The plots suggest that the "seep through" to the tail is strongly dependant on the wavefunction's orientation.*

- $z_{imp} = 50nm$ here the impact of the impurity position has been pronounced with the electron having the highest probability of being localized around the impurity position on the boundary between the head and the tail;
- $z_{imp} = 100nm$ here the electron can be described to be in an intermediate state between being localized in the head and being localized in the tail;
- $z_{imp} = 150nm$ and finally, the electron is almost entirely localized in the tail with a small probability of being in the head, in stark contrast to the ground state where it was too far away to "feel" the impurity.

The effects described above have a strong dependence on the orientation of the PD, which is demonstrated in Figure 3 where the PDs of the first four energy levels are given for the impurity position $z_{imp} = 0$. The first level was discussed in detail, that is why we will skip the discussion. The first three excited states are similar to the spherical QD, in wich they are degenerate. The difference between these three states is the orientation: the $n = 2$ is oriented with respect to the z-axis $n = 3, 4$ are oriented to the x and y-axis. When the wave function is oriented to the x and y-axis it is fully localized in the head region. However, the z orientation leads to the partial elimination of the degeneracy allowing the PD to "seep through" to the tail. As a result the energy of this state will be different from the $n = 3, 4$ cases, which can be seen in the next figure.

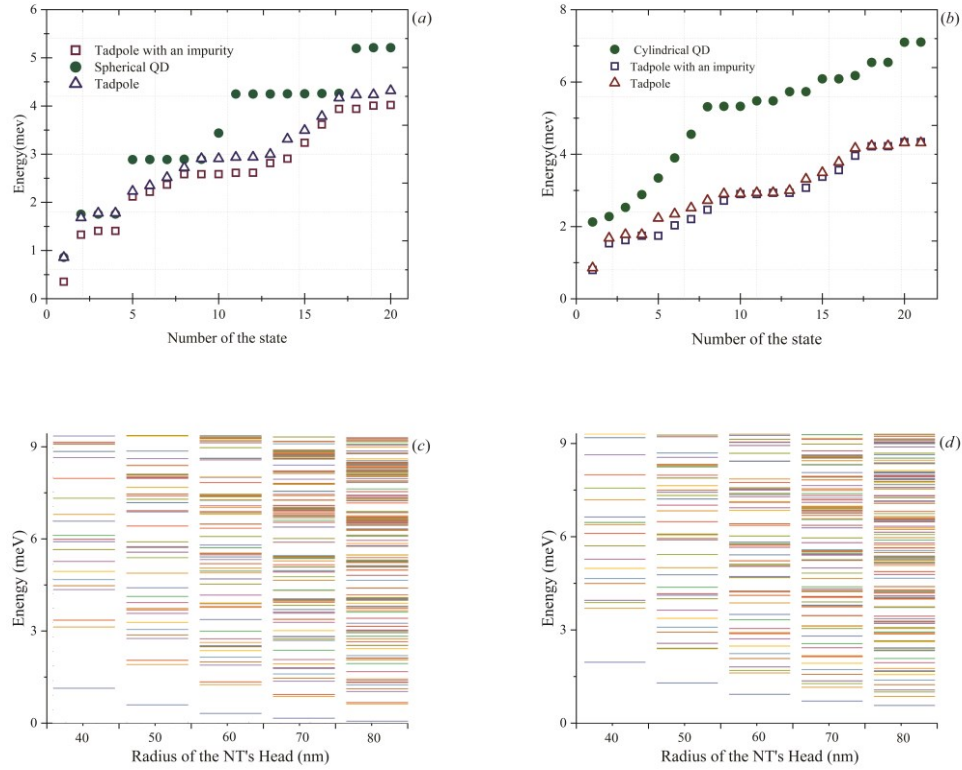

**Figure 4.** *(a) The first 20 energy levels for the cylindrical quantum dot, chalcopyrite AgInSe$_2$ nanotadpole without an impurity, nanotadpole with an impurity $z_{imp} = 150nm$. (b) spherical quantum dot, nanotadpole without an impurity, nanotadpole with an impurity $z_{imp} = 0$. It was observed that NT's energy levels mostly exhibit behaviors closer to the spherical QD(a) which is evident from the large energy gap between the nanotadpole's energy levels and the cylindrical quantum dot (b). In the cases where the wave function's orientation and shape allow the electron to be mostly localized in the tail, the energy behaves more like a cylinder. In the remainder of cases, it exhibits the behavior of a hybrid state between those two symmetries. (c) The energy spectrum with the impurity position $z_{imp} = 0$. (d) The energy spectrum with the impurity position $z_{imp} = 150nm$.*

The next logical step would be the discussion of the energy levels of the chalcopyrite AgInSe2 NTs. The analysis of the PD implies that depending on the position of the impurity and the orientation of the wave function the energy levels can be predicted to behave like either a spherical QD, a cylindrical QD, or a hybrid symmetry between those two. The comparison between the spherical QD's, NT's, and NT's with impurity energy levels depending on the number of the state, is presented in Figure 4 (a) and (b) for two different positions of impurity. Reviewing the results outlined in Figures 4(a,b), it can be concluded that the NT's energy levels' behavior is mostly analogous to that of a spherical QD, in a sense that it has clearly defined degenerate states, however, unlike the spherical QD in a range those degenerate states the energy levels show more cylindrical behavior. In both cases interaction of the electron with the impurity leads to the decrease of the energy levels. The energy spectrum for the NT's different sizes with two different impurity positions has been obtained and showcased in Figure 5(c,d) – $R_{head}=40nm, 50nm, 60nm, 70nm, 80nm$, $R_{tail}=R_{head}/2$. Where it is visible that with the decrease of the NT's size the energy levels rise and the difference between energy levels $\Delta E_{ij}=E_i-E_j$ increases. A detailed investigation of the spectra is required for future calculations.

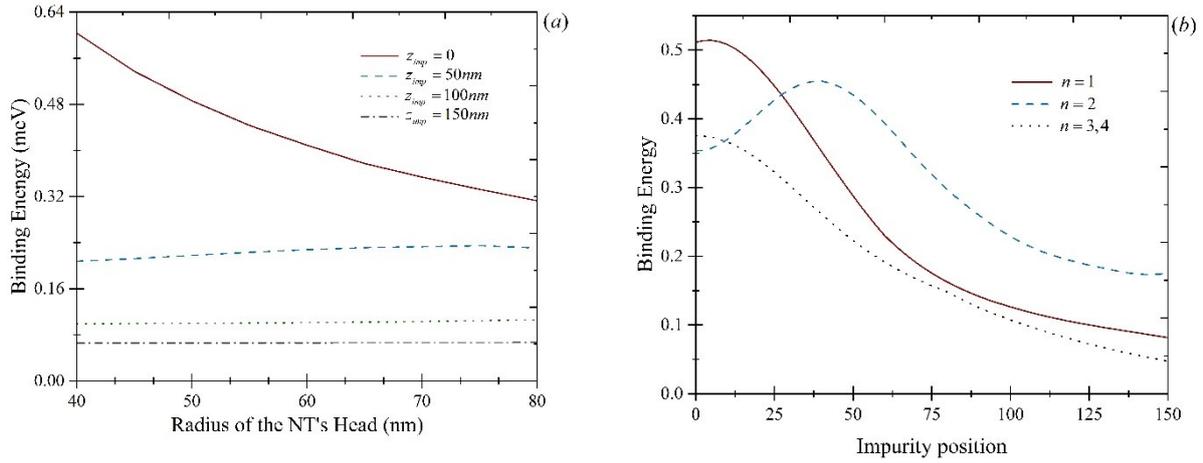

**Figure 5.** *The binding energy's dependence on the chalcopyrite AgInSe$_2$ nanotadpole's size: changing the head's radius from $40nm$ to $80nm$ while the radius of the tail is kept to be the half of the head's, for different impurity positions(a). The figure clearly shows that the binding energy $z_{imp}=0$ is the most susceptible to the change of the nanotadpole's size i.e. the binding energy decreases with the increase of the size. The case of the $z_{imp}=50nm$ binding energy shows a slow increase with the increase of the size in the discussed range. The last two cases $z_{imp}=100nm, 150nm$ practically show no response to the change of size. The binding energy's dependence on the impurity position for the first four states is presented in (b). The figure shows a clearly defined maximum around $z_{imp}=0$ for the $n=1,3,4$ cases. For the first excited state, the maximum is shifted to the right.*

The binding energy which is defined as $E_{bind}=E^i-E^i_{imp}$ where $E^i$ is the energy of the $i^{th}$ state without the impurity and $E^i_{imp}$ represents the energy of the same state with the impurity, is an important property of structures with hydrogenic impurities. Figure 5a represents the dependence of the binding energy on the NTs size for the different positions of the impurity. The thorough investigation of the binding energy shows that for the ground state:

- $z_{imp}=0$ the binding energy has an inverse dependence on the size of the NT and shows the highest responsiveness. With the increase of NT's size, the localization region of the electron increases, which negatively affects the interaction between electron and impurity center.

- $z_{imp}=50nm$ has a direct dependence on the size, in the investigated range. The same localization effect brings positive effects on the interaction between electron and impurity center; thus, the binding energy increases slightly in the investigated range.

- $z_{imp}=100nm, 150nm$ show no observable dependence in the investigated range, as the change in the localization area, does not affect the interaction between far impurity and electron PD.

Figure 5b presents the dependence of the binding energy on the impurity positions for different states. With the shift of impurity position the binding energy of the ground state increases slightly reaches its maximum around the point $z_{imp} \approx 5nm$ and then decreases. The slight increase is explained by the slight asymmetric character of PD and the further decrease is explained by the impurity leaving from the spherical region i.e., the electron's localization region. The binding energy for the first excited state displays an interesting behavior: when the impurity is at the center it is noticeably slower than the binding energy for the ground state, parallel to the shift of $z_{imp}$ it starts to increase, reaching the maximum point $z_{imp} \approx 40nm$ after which it again starts to decrease. The existence of the maximum point can be explained by the help of the second case for the excited state presented in Figure 2b. Discussing it further, in the case when we have spherical QD, the binding energy has a maximum value when the impurity is localized in the center of the dot. While in the NT's case, the PD changes as the impurity position come closer to the boundary between the sphere and the cylinder, the confinement effects increase, increasing the interaction between the impurity and the electron. The binding energy for the second and third excited states has similar behavior to the ground state. Starting from a certain value, they decrease parallel to the impurity shift to the tail. As it is expected the binding energies for these states completely match each other.

## 3.1 Diamagnetic Susceptibility

A thorough investigation of the chalcopyrite AgInSe$_2$ NT's DS's dependency on the size and impurity position was carried out. The review of these results suggests that the DS's dependence on the size is virtually constant for the investigated range. It is important to note, that during the investigative process the morphology was not changed, meaning the relation of the head's size and the tail's size were kept constant.

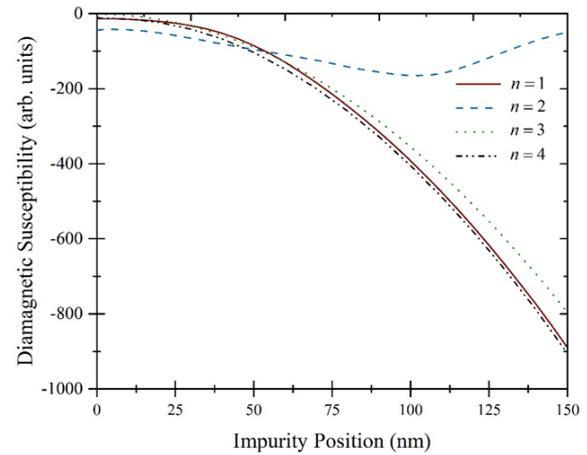

**Figure 6.** *The dependence of the diamagnetic susceptibility on the impurity position $z_{imp}$ for the first four states of the AgInSe$_2$ nanotadpole is presented. The analysis of the curves implies that for the $n = 1, 3, 4$ state the diamagnetic susceptibility decreases starting at their maximum around the $z_{imp} = 0nm$ point. The $n = 2$ decrease was much slower than the other states reaching a local minimum and increasing until reaching saturation.*

During the investigative process, it was found the NT's DS is strongly dependent on the impurity position. The DS's dependence on the impurity position is given in Figure 6 for the first four energy levels. The results showcase that for the ground state, second ard third excited states the DS decreases from the $z_{imp} = 0nm$ point, while for the first excited state the DS decreases until reaching a local minimum. After which it starts to rise and predictably becomes saturated as the impurity goes further into the tail. This effect can be explained by the transition of the electron's probability density from the spherical symmetry to the cyllindrical which is clearly shown in the Figure 2.

## 3.3 *Photoionization Cross Section*

The PCS of the chalcopyrite AgInSe2 NTs have been calculated for $i = 1, 2, 3$ initial states and $f = 2, 3, ..., 21$ final states. Because of the strong dependence of the matrix element on the impurity position we have considered the four most interesting impurity positions for the PCS calculations. The incident angle is chosen to be $\alpha = 0$ with respect to z-axis which corresponds to the maximum of the matrix element. It is also important to note that the population of the levels neglected for these calculations.

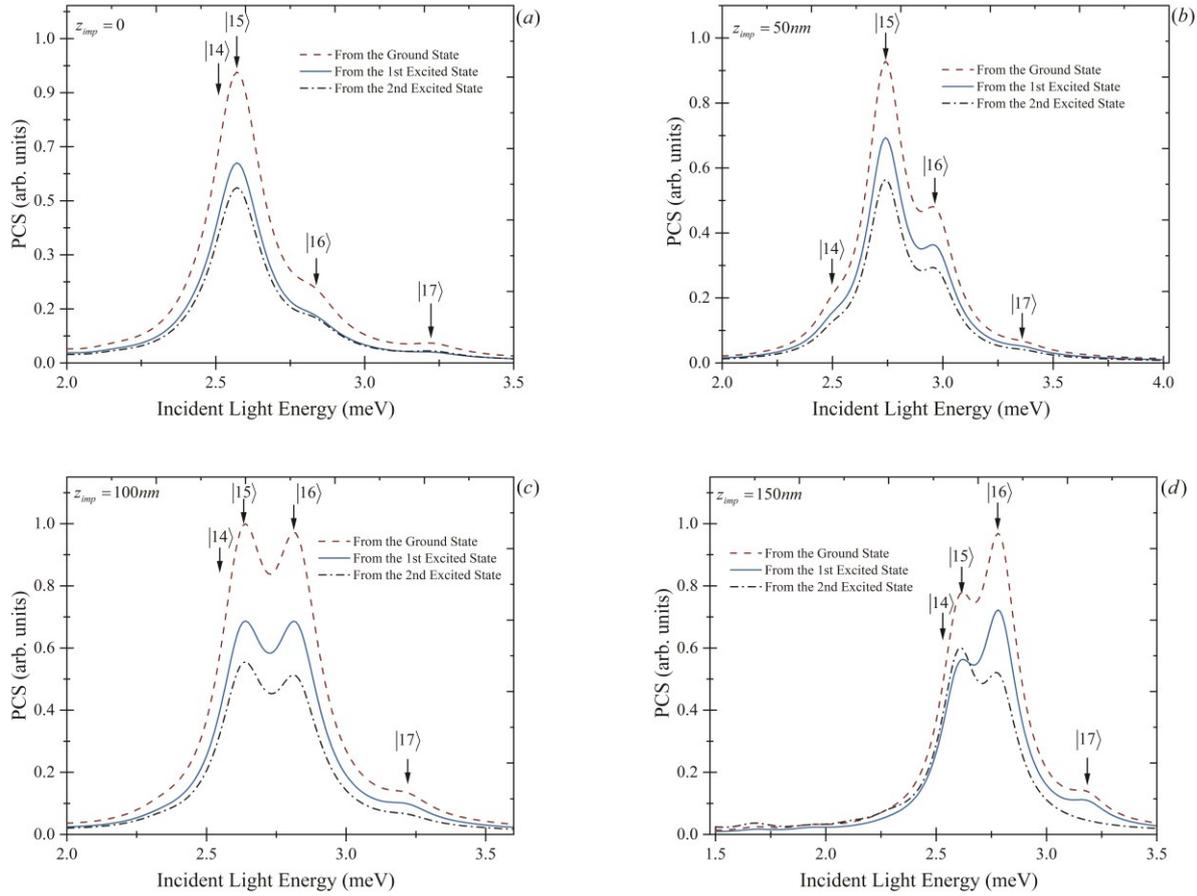

**Figure 7.** *The photoionization cross-section for four different impurity positions $z_{imp} = 0nm$, $50nm$, $100nm$, $150nm$ for chalcopyrite AgInSe$_2$ nanotadpole is presented. The four transitions that have the highest contribution to the PCS $|i\rangle \to |14\rangle$, $|i\rangle \to |15\rangle$, $|i\rangle \to |16\rangle$, $|i\rangle \to |17\rangle$, $i = 1, 2, 3$ their corresponding peaks are presented in the figure. The transitions are calculated from the presented states (ground – dashed line, 1$^{st}$ excited - blue solid line, 2$^{nd}$ excited – black dashed-dotted line) to the final state and the transitions are depicted only by the final state. The results show that the $z_{imp} = 0nm$ case transitions to the $|15\rangle$ state result in the highest peak (a). As the impurity position shifts to $z_{imp} = 50nm$ the peak of the transition to the $|16\rangle$ state have a higher magnitude compared to $z_{imp} = 0nm$ case (b), while the $|15\rangle$ remains the dominant peak. In this $z_{imp} = 100nm$ case, the photoionization cross-section has two peaks corresponding to transitions to the $|15\rangle$ and $|16\rangle$. The peak corresponding to the $|15\rangle$ remains marginally higher. Finally, for the $z_{imp} = 150nm$ case, the transitions to the $|16\rangle$ state overtake the dominance from $|15\rangle$ for all cases but $i = 2$.*

The review of the results shows that the transitions with the highest contributions are the transitions from the initial states to the $|14\rangle, |15\rangle, |16\rangle, |17\rangle$ which are shown in Figure 7. The PCS curve radically changes with the change of the impurity position. When the impurity is situated at the center of the head $z_{imp} = 0nm$ figure 7a the peaks for the transitions to the $|14\rangle$ and $|15\rangle$ states are merged and corresponds to the highest maximum, the transitions to the $|16\rangle$ and $|17\rangle$ correspond to two separate peaks with lower intensities. When the impurity's position is at $z_{imp} = 50nm$ the figure 7c the peaks for the transitions to the $|14\rangle$ and $|15\rangle$ states are split and the peak position of the $|14\rangle$ is slightly visible. The transition to the $|15\rangle$ still has the highest peak intensity and the individual PCS peak for the $|16\rangle$ state is much more discernable. The next case is where the impurity position is in the tail region and $z_{imp} = 100nm$ the peaks for $|14\rangle$ and $|15\rangle$ transitions merge and result in a marginally higher peak compared to the $|16\rangle$ transition. The intensity of the $|17\rangle$ relative to the other peaks remains unchanged. Finally, for the $z_{imp} = 50nm$ case the figure 7d although the peaks of the $|14\rangle$ and $|15\rangle$ transitions are merged the highest peak corresponds to the $|16\rangle$ transition in all cases except for $i = 3$.

## 3. Conclusion

In conclusion, the optical and magnetic properties of the AgInSe$_2$ NT's were investigated. The PD of the electron was investigated and it was found that because of the asymmetrical nature of the NT's morphology the PD showed a strong dependence on the impurities position and on the orientation of the wave function. For the ground state, the electron largely remains localized in the head. And for the excited states, where the wavefunction is oriented towards the z-axis, the electron has a considerable "seep through" and the impurity position has a large contribution to the shape and the localization of the PD. The energy spectrum for different impurity positions has been investigated. The review of the results shows that the NT's energy spectrum largely behaves similar to the spherical QD, except the cases where the wavefunction's orientation and the impurity position allow the electron to be mostly localized in the tail which allows the PD to behave closer to the cylindrical QD case.

The binding energy dependent on the NT's size and the impurity position was obtained. The binding energy slowly increases with the decrease of the NT's size for all cases but $z_{imp} = 0nm$, where it sharply decreases. The results for the binding energy dependent on the impurity position show a clearly defined maximum around $z_{imp} = 0nm$ for the $n = 1, 3, 4$ cases. For the first excited state, the maximum is shifted to the right.

The dependence of the DS on the impurity position $z_{imp}$ for the first four states is obtained. The review of the results implies that for the $n = 1, 3, 4$ state the DS decreases starting at their maximum around the $z_{imp} = 0nm$ point. For the $n = 2$ case decrease was much slower than the other states reaching a local minimum and increasing until reaching saturation which is explained by the bahaviour of the PD. Finally, the PCS was calculated for the $i = 1, 2, 3$ initial states and $f = 2, 3, ..., 21$ final states with different impurity positions. The individual PCS with the highest contribution were found to be the transitions from $i = 1, 2, 3$ to $|14\rangle, |15\rangle, |16\rangle, |17\rangle$. The analysis of the data shows that the $z_{imp} = 0nm, 50nm$ lower transitions $|14\rangle, |15\rangle$ result in the highest peak and higher transitions become more prevalent $|16\rangle, |17\rangle$ when the impurity shifts further into the tail. In the case of $z_{imp} = 150nm$ the peak for $|16\rangle$ has the highest intensity.

## Appendix

The matrix element calculations for the first 9 transitions starting from the ground state are given in Table 1 for three different incident angles and four different impurity positions. Here it is visible that for the considered transitions the matrix element has taken the maximum value when the incident angle $\alpha = 0$ for all impurity positions. Moreover, in the case of the $\alpha = \pi/2$ transitions are not allowed. The review of the obtained data also suggests that the impurity position has a major contribution to the matrix element in some cases where the transitions are not allowed for the $z_{imp} = 0$ become allowed when the impurity shifts further. The same patterns that are visible for the Table 1 are visible for transitions from the first excited state. The transitions starting from the first excited state are given in Table 2.

**Table 1:** *The matrix element for transitions from the ground state to the first nine excited states.*

| $z_{imp}(nm)$ | $\alpha$ | $1\to 2$ | $1\to 3$ | $1\to 4$ | $1\to 5$ | $1\to 6$ | $1\to 7$ | $1\to 8$ | $1\to 9$ | $1\to 10$ |
|---|---|---|---|---|---|---|---|---|---|---|
| 0 | 0 | 0.056 | 0 | 0 | 0.106 | 0.067 | 0.174 | 0.2 | 0 | 0 |
| 50 | $\pi/4$ | 0.028 | 0 | 0 | 0.053 | 0.033 | 0.087 | 0.100 | 0 | 0 |
|  | $\pi/2$ | 0 | 0 | 0 | 0 | 0 | 0 | 0 | 0 | 0 |
|  | 0 | 0.951 | 0 | 0 | 0.613 | 0.023 | 0.171 | 0.120 | 0 | 0 |
| 100 | $\pi/4$ | 0.475 | 0 | 0 | 0.306 | 0.011 | 0.085 | 0.060 | 0 | 0 |
|  | $\pi/2$ | 0 | 0 | 0 | 0 | 0 | 0 | 0 | 0 | 0 |
|  | 0 | 0.429 | 0.009 | 0 | 0 | 0.209 | 0.037 | 0.120 | 0.026 | 0 |
| 150 | $\pi/4$ | 0.214 | 0.004 | 0 | 0 | 0.104 | 0.018 | 0.060 | 0.013 | 0 |
|  | $\pi/2$ | 0 | 0 | 0 | 0 | 0 | 0 | 0 | 0 | 0 |
|  | 0 | 0.033 | 0.110 | 0 | 0 | 0.172 | 0.045 | 0.192 | 0.063 | 0 |
| $z_{imp}(nm)$ | $\pi/4$ | 0.016 | 0.055 | 0 | 0 | 0.086 | 0.022 | 0.096 | 0.031 | 0 |
|  | $\pi/2$ | 0 | 0 | 0 | 0 | 0 | 0 | 0 | 0 | 0 |
|  | $\alpha$ | $1\to 2$ | $1\to 3$ | $1\to 4$ | $1\to 5$ | $1\to 6$ | $1\to 7$ | $1\to 8$ | $1\to 9$ | $1\to 10$ |

**Table 2:** *The matrix element for transitions from the excited state to the first nine excited states.*

| $z_{imp}(nm)$ | $\alpha$ | $2\to 3$ | $2\to 4$ | $2\to 5$ | $2\to 6$ | $2\to 7$ | $2\to 8$ | $2\to 9$ | $2\to 10$ |
|---|---|---|---|---|---|---|---|---|---|
| 0 | 0 | 0 | 0 | 1.15 | 0.026 | 0.813 | 0.181 | 0 | 0 |
|  | $\pi/4$ | 0 | 0 | 0.575 | 0.013 | 0.406 | 0.09 | 0 | 0 |
|  | $\pi/2$ | 0 | 0 | 0 | 0 | 0 | 0 | 0 | 0 |
| 50 | 0 | 0 | 0 | 5.239 | 0.241 | 0.293 | 0 | 0 | 0 |
|  | $\pi/4$ | 0 | 0 | 2.619 | 0.12 | 0.146 | 0 | 0 | 0 |
|  | $\pi/2$ | 0 | 0 | 0 | 0 | 0 | 0 | 0 | 0 |
| 100 | 0 | 42.55 | 0 | 0 | 0.271 | 1.296 | 1.211 | 0.027 | 0 |
|     | $\pi/4$ | 21.27 | 0 | 0    | 0.135 | 0.648  | 0.605 | 0.013 | 0 |
|     | $\pi/2$ | 0     | 0 | 0    | 0     | 0      | 0     | 0     | 0 |
| 150 | 0     | 18.94 | 0 | 0.06 | 1.628 | 16.219 | 0     | 1.263 | 0 |
|     | $\pi/4$ | 9.47  | 0 | 0.03 | 0.814 | 8.109  | 0     | 0.631 | 0 |
|     | $\pi/2$ | 0     | 0 | 0    | 0     | 0      | 0     | 0     | 0 |